\documentstyle[12pt]{article}
\title{
{}\hfill
    \raisebox{2cm}[0pt]{
     \begin{array}[b]{l}
       \small\mbox{HUTP-A054/98, ITEP-TH-62/98, }
\small\tt
hep-th/9810147
  \end{array}
    }\\
  On noncommutative Nahm transform.}
\author{Alexander Astashkevich$^{1}$, Nikita
Nekrasov$^{2}$ and
Albert Schwarz$^{1}$\\
\\
\footnote{ast@math.ucdavis.edu, nikita@string.harvard.edu,
schwarz@math.ucdavis.edu}\\
\\
$^{1}$ Department of Mathematics,
        UC Davis\\
$^{2}$ Lyman Laboratory of Physics,
	Harvard University\\
$^{2}$ Institute of Theoretical and Experimental Physics,
Moscow }

\newcommand\Z{\rm Z\!\!Z}
\newcommand{\C}{\rm I\!\!\!C}
\newcommand{\R}{\rm I\!R}

\newtheorem{thm}{Theorem}[section]
\newtheorem{lem}{Lemma}[section]

\newtheorem{prop}{Proposition}[section]

\newtheorem{defi}{Definition}[section]

\newtheorem{rem}{Remark}[section]
\newtheorem{claim}{Claim}[section]

\def\At{{{\cal A}_{\theta}}}
\def\Aht{{{\cal A}_{\hat{\theta}}}}
\def\Adt{{{\cal A}_{\theta\oplus(-\hat{\theta})}}}
\def\P{{\cal P}}
\def\CR{{\nabla}^R}
\def\ChR{{\nabla}^{\hat{R}}}
\def\n{\nabla}
\def\RtP{{R\otimes_{\At}\P}}
\def\Lt{{L_{\theta}}}
\def\Lht{{L_{\hat{\theta}}}}
\def\N{{\cal N}}
\def\A{\cal A}
\def\Id{{\mbox{\rm Id}}}

\begin{document}
\maketitle

Motivated by the recently observed relation between the
physics of $D$-branes in the
background of $B$-field and the noncommutative geometry
we study the analogue
of Nahm transform  for the instantons on the
noncommutative torus.

\section{Introduction}
\label{sec1}

It is shown in \cite{cds} that noncommutative geometry can be
successfully
applied to the analysis of M(atrix) theory. In
particular, it is
proven
that one can compactify M(atrix) theory on a
noncommutative torus;
later, compactifications of this kind were studied in numerous
papers.
In present paper, we analyze instantons in noncommutative
toroidal
compactifications. This question is important because
instantons can
be
considered as BPS states of compactified M(atrix) model.
Instantons
on a noncommutative $\R^4$ were considered earlier in
\cite{ns}. It
is shown
in \cite{ns} that these instantons give some insight in
the structure
of
(2,0) super-conformal six dimensional theory; the
instantons on a
noncommutative
torus also should be useful in this relation. The main
mathematical tool used in \cite{ns} is the noncommutative
analogue of
ADHM
construction of instantons. The present paper is devoted
to the
noncommutative analogue of Nahm transform (recall that
the Nahm
transform
can be regarded as some kind of generalization of ADHM
construction).
We prove that some of important properties of Nahm
transform remain
correct in
noncommutative case.

\section{Preliminaries.}
\label{sec2}

In this section we  recall several notions related to the
theory of
noncommutative tori. We roughly discuss the ideas behind the
noncommutative
Nahm transform and formulate our main results. A more formal
approach to
the
noncommutative Nahm
transform
is taken in the next section.

\begin{defi}
\label{tor}
An $n$-dimensional noncommutative torus ${\cal A}_{\theta}$
is a $C^*$-algebra having unitary generators
$U_{\bf i}$, ${\bf i}\in\Z^n$ obeying
\begin {equation}
U_{\bf i}U_{\bf j}=e^{\frac{i}{2}\theta({\bf i,
j})}U_{\bf i+j},
{}~~~~~~~
U_{\bf 0}=1;
\end {equation}
where $\theta({\bf \cdot , \cdot})$ is a skew-symmetric
bilinear
form on $\Z^n$.
\end{defi}
We can naturally consider
$\theta$ as a skew-symmetric bilinear form on ${\R}^n$.
Any element of  $\At$ can be uniquely  represented   as a sum
$\sum_{{\bf i}\in\Z^n} c_{\bf i}U_{\bf i}$,
where $c_{\bf i}$ are complex numbers.
Let $e_k$, $1\leq k\leq n$
be the natural base in $\Z^n$. The transformations $\delta _l
U_{e_k}=\delta _{l,k}U_{e_k}$, $1\leq k,l\leq n$ generate
an abelian
Lie algebra $L_{\theta}$ of
infinitesimal automorphismes of ${\cal A}_{\theta}$. We  use
$L_{\theta}$ to define the notion of connection in a ${\cal
A}_{\theta}$-module following \cite{Con1} (we do not need
the general
notion of connection \cite{Con2}).

Any element from $\At$ can be considered as a function
on the $n$-dimensional torus whose Fourier coefficients
are $c_{\bf
i}$
(see above).  The space of smooth functions on $T^n$ forms a
subalgebra
of $\At$. We denote it by ${\At}^{smooth}$ and call it
the smooth
part of
$\At$. If $E$ is a projective finitely generated $\At$
module one can
define its smooth part $E^{smooth}$ in a similar manner (see
\cite{Rf1}).
Now we can define the notion of $\At$ connection.

\begin{defi}
\label{def0}
{\it $\At$ connection on a right $\At$ module $E$
is a linear map $\nabla$ from
$L_{\theta}$ to the space $\mbox{\rm End}_{\At}E$ of
endomorphisms of $E^{smooth}$, satisfying
the condition
$$\n_{\delta}(ea)=(\n_{\delta}e)a+e(\delta(a)),$$
where $e\in E^{smooth}$, $a\in\At^{smooth}$, and $\delta\in
L_{\theta}$.
The curvature $F_{\mu,\nu}=[\n_{\mu},\n_{\nu}]$ of
connection $\n$ is
considered as a two-form on $\Lt$ with values in
endomorphisms of
$E$.}
\end{defi}

We always consider Hermitian modules and Hermitian
connections.
This means that if $E$ is a right $\At$ module it is
equipped with
$\At$ valued Hermitian inner product
$\langle\cdot,\cdot\rangle$
(for the detailed list of properties see \cite{Bl}) and all
connections
that we will consider should be compatible with this
inner product.

If $E$ is endowed with a $\At$-connection, then one can
define a
Chern character
\begin{equation}
\label{eqchern}
\mbox{\rm ch}(E)=\sum_{k=0}\frac{\hat{\tau}(F^k)}{(2\pi
i)^k k!},
\end{equation}
where $F$ is a curvature of a connection on $E$, and
$\hat{\tau}$ is
the
canonical trace on $\hat{A}=\mbox{\rm End}_{\At}(E)$
(we use that
 $\At$ is equipped with a canonical trace $\tau=c_0$). One can
consider
 $\mbox{\rm ch}(E)$ as an element in the  Grassmann algebra
$\Lambda(\Lt^*)$.
However, it is convenient to use the fact that this
Grassmann algebra
can be identified with cohomology $H( T,\C)$ where $ T$
stands for
the
Lie group of automorphisms of the algebra $\At$
corresponding to the Lie algebra
$L_\theta$. (In other words $T=L_\theta / D$ where $D$ is
a lattice.)
In the commutative case ${\rm ch}(E)$ is an integral
cohomology
class.
In noncommutative case this is wrong, but there exists an
integral
cohomology class $\mu (E) $ $\in H(T,\Z)$ related to
 $\mbox{\rm ch}(E)$  by the
following formula (see \cite{Ell}, \cite{Rf1})
\begin{equation}
\label{eqrelchmu}
\mbox{\rm ch}(E)=e^{\iota(\theta)}\mu(E),
\end{equation}
Here ${\iota(\theta)}$ stands for the operation of
contraction with
${\theta}$ considered as an element of two-dimensional
homology group
of $T$.
In particular, formula (\ref{eqrelchmu}) means that
$e^{-\iota(\theta)}\mbox{\rm ch}(E)\in\ H(T,\Z)$.

One can regard $\mu(E)$ as a collection of integer
quantum numbers
characterizing topological class of a gauge field on
noncommutative
torus (or from mathematical viewpoint as a K-theory class of
projective
module $E $.)

The formula (\ref{eqrelchmu}) is familiar to physicists
\cite{D-WZ}
in the following ($T$-dual) form:
\begin{equation}
\label{phys}
{\cal L}_{WZ}= \int_{X} \nu\wedge \tilde C, \,\,\  \nu =
\mbox{\rm
ch}(E)
e^{- \frac{B}{2\pi i}}
\sqrt{{\hat A}(X)}
\end{equation}
The element $\nu$ of the cohomology group $H^{\rm
even}(X, {\R})$ is
called the
generalized Mukai vector.
Here ${\cal L}_{WZ}$ describes the so-called Wess-Zumino
couplings
on the worldvolume $X$ of a
$D$-brane (more precisely of a stack of $D$-branes), $E$
is the
Chan-Paton bundle (or
sheaf), $\tilde C$ is the collection of all Ramond-Ramond
potentials,
${\hat A}(X)$ is the A-roof genus  and $B$ is the
Neveu-Schwarz
$B$-field. The formula (\ref{phys}) captures the effect of the
non-trivial topology of the Chan-Paton bundle on the $D$-brane
charges induced on the brane.
Since in our case $X$ is a torus with flat metric then
${\hat A}(X) =
1$ and we arrive at
(\ref{eqrelchmu}) provided that we performed a $T$-duality
transformation which maps $B$-field
two form into a bivector $\theta$ and also exchanges
${\rm ch}_{k}$
with ${\rm ch}_{\frac{n}{2}-k}$.

\begin{defi}
\label{inst}
An instanton is a connection such that the
self-dual part
of
its curvature is a scalar operator,  i.e.,
$F^+_{\alpha\beta}$ is a multiplication operator by the
scalar that we denote $-\omega^+_{\alpha\beta}$.
\end{defi}

We are interested in instantons on four-dimensional
noncommutative
torus.
In the framework of supersymmetric gauge theory they can be
interpreted
as BPS-fields. Notice that in the definition of Hodge
dual $\star F$ we
need
an inner product on the Lie algebra $\Lt$; we fix such a
product.

As in commutative case we can prove that the minimum of
euclidean
action
$A(E,\nabla)=\frac{1}{8\pi^2} \hat{\tau}(F\wedge \star F)[T]$
for  connections  in module $E$ (i.e. for gauge fields
with given
topological
numbers) is achieved on instantons or anti-instantons
(connections
where antiselfdual part of curvature is a scalar).
Here $[T]$ stands for the fundamental homology class of $T$.

\begin{claim}
\label{action}
The expression
of the instanton action in terms of $\mu$ and $\theta$ has the
following form :
\begin{equation}
\label{eqforact}
A(E,\nabla)=\left(\mu_2(E)-
\frac{i}{2\pi }
\frac{\left(\mu_1(E)^{+}+\left(
\iota(\theta)\mu_2(E)\right)^{+}\right)^{2}}{\mu_{0}(E) +
\iota(\theta)\mu_1(E) +
\frac{1}{2} \iota(\theta)^2 \mu_2(E)} \right)[T],
\end{equation}
where $\mu_k(E)$ is $2k$-dimensional component of $\mu(E)$.
\end{claim}
To prove this formula we notice
that if $(E,\nabla)$ is an instanton ({\it i.e.}
$F^++\omega^+=0$)
the we can express $*F$ as follows
\begin{equation}
\label{eqinst}
\star F=-F-2\omega^+.
\end{equation}
Using (\ref{eqinst}) we obtain
$$\hat{\tau}(F\wedge \star F)=\hat{\tau}(F\wedge
[-F-2\omega^+])=
-\hat{\tau}(F\wedge F)-2\hat{\tau}(F)\wedge\omega^+,$$
since $\omega$ is complex valued two-form  hence
$$
\hat{\tau}(F\wedge\omega^+)=\hat{\tau}(F)\wedge\omega^+ = -
\hat{\tau}(1) \omega^+ \wedge \omega^+.
$$
One can easily obtain  from
the formula (\ref{eqrelchmu}) that
$\hat{\tau}(F\wedge F)=-8\pi^2\mu_2(E)$,
$\hat{\tau}(F)=2\pi i (\mu_1(E)+\iota(\theta)\mu_2(E))$.
$\Box$.

We will construct a
generalization of Nahm's transform \cite{Na}, \cite{Dk}
relating connections on ${\cal
A}_{\theta}$-modules with connections on ${\cal
A}_{\hat{\theta}}$-modules.
(Here ${\cal A}_ {\theta}$ and ${\cal A}_{\hat
{\theta}}$ are two four-dimensional noncommutative tori.)
To define a noncommutative generalization of Nahm's
transform we need
a
$({\cal A}_{\theta}, {\cal A}_{\hat {\theta}})$-module
${\cal P}$
with
${\cal A}_{\theta}$-connection $\nabla$ and ${\cal A}_{\hat
{\theta}}$-connection $\hat{\nabla}$. Speaking about $({\cal
A}_{\theta},{\cal A}_{\hat {\theta}})$-module ${\cal P}$
we have in
mind
that ${\cal P}$ is a left ${\cal A}_{\theta}$-module and
a right
${\cal
A}_{\hat {\theta}}$-module; and $(ax)b=a(xb)$ for $a\in {\cal
A}_{\theta},\  b\in {\cal A}_{\hat {\theta}},\  x\in
{\cal P}$.
We assume that the
commutators $[\nabla _{\alpha}, \nabla _{\beta}],\  [\hat
{\nabla}_{\mu},
\hat {\nabla}_{\nu}],\  [\nabla _{\alpha},\hat
{\nabla}_{\mu}]$ are
scalars:
$$[\nabla _{\alpha}, \nabla _{\beta}]=\omega
_{\alpha\beta},\  [\hat
{\nabla}_{\mu}, \hat {\nabla}_{\nu}]=\hat {\omega}_{\mu\nu},\
[\nabla
_{\alpha},\hat {\nabla}_{\mu}]=\sigma _{\alpha\mu}.$$
One more assumption is that $\nabla _{\alpha}$ commutes
with the
multiplication by the elements of ${\cal A}_{\hat {\theta
}}$ and
$\hat
{\nabla} _{\alpha}$ commutes with the multiplication by
the elements
of ${\cal
A}_{\theta }$.
One can reformulate the above conditions saying that
$\P$ is an  ${\cal A}_{\theta\oplus(-\hat{\theta})}$-module,
and $\nabla,\hat{\nabla}$ give us a constant curvature
connection on
it.
For every  right ${\cal A}_{\theta}$-module $R$ with
connection
$\nabla^R$
we consider Dirac operator $D=\Gamma ^{\alpha}(\nabla
_{\alpha}^R+\nabla _{\alpha})$ acting on the tensor product
$$(R\otimes_{{\cal A}_{\theta}}{\cal P})\otimes S$$
(or more precisely on its smooth part).
To define $\Gamma$-matrices we introduce an inner product in
$L_{\theta}$.
$S$ is a $\Z/2\Z$ graded vector space $S=S^{+}\oplus S^-$ and
Dirac operator is an odd operator. Thus, we can consider
\begin{equation}
\label{eq10}
D^+:(R\otimes_{{\cal A}_{\theta}}{\cal P})\otimes
S^+\rightarrow
(R\otimes_{{\cal A}_{\theta}}{\cal P})\otimes S^-,\\
D^-:(R\otimes_{{\cal A}_{\theta}}{\cal P})\otimes
S^-\rightarrow
(R\otimes_{{\cal A}_{\theta}}{\cal P})\otimes S^+.
\end{equation}

The Dirac operator commutes with the multiplication by
the elements
of ${\cal
A}_{\hat {\theta }}$, therefore the space of zero modes
of $D$ can be
regarded as $\Z/2\Z$ graded ${\cal A}_{\hat {\theta}}$-module;
we denote it by ${\hat{R}}$.
We will see later that sometimes it is reasonable to
modify this
definition
of $\hat{R}$.

Next we would like to get a connection on $\hat{R}$.
Using $\At$ Hermitian inner product on $R$,
$\At\times\Aht$ Hermitian
inner product on $\P$, and a canonical trace $\tau$ on $\At$
we can define an $\Aht$ Hermitian
inner product on $(R\otimes_{{\cal A}_{\theta}}{\cal
P})\otimes S$.
We assume that we have an orthogonal projection
(with respect to the $\Aht$ Hermitian inner product)
$P$  of $(R\otimes_{{\cal A}_{\theta}}{\cal P})\otimes S$ onto
$\hat{R}$.
We will prove its existence in the next section under
some additional
conditions.
The connection $\hat {\nabla}$ induces a connection $\hat
{\nabla}^{\prime}$ on
$$(R\otimes _{{\cal A}_{\theta}}{\cal P})\otimes S;$$
we obtain a connection $\nabla^{\hat{R}}$ on $\hat {R}$
as $P\hat
{\nabla}^{\prime}$.

The above construction can be regarded as a generalized Nahm's
transform. To
prove that its properties are similar to the properties
of standard
Nahm's
transform we should impose additional conditions on
module ${\cal P}$
and
connections $\nabla,\, \hat {\nabla}$. We assume that
$\cal P$ is a projective Hermitian
module over ${\cal A}_{\theta\oplus{(-\hat{\theta})}}$
and that $\sigma _{\alpha \mu}$ determines a
non-degenerate pairing
between
$L_{\theta}$ and
 $L_{\hat {\theta}}$.  Then we can use this pairing to
define an
inner
product in
 $L_{\hat{\theta}}$;  this allows us to talk about
instantons on
$\Aht$. Under certain conditions we prove that the Nahm
transform of
an
instanton is again an instanton.
More precisely, if
$F^+_{\alpha\beta}+\omega^+_{\alpha\beta}=0$
where $\omega^+_{\alpha\beta}$ stands for the selfdual
part of
$\omega_{\alpha\beta}$, then $\hat{F}^+_{\alpha\beta}-
\hat{\omega}^+_{\alpha\beta}=0$. Here $F_{\alpha\beta}$
(correspondingly
$\hat{F}_{\alpha\beta}$)  is the curvature
of the connection $\n^R$ (correspondingly
$\nabla^{\hat{R}}$) on $R$
(correspondingly $\hat{R}$)  and $^+$ stands for the
self-dual part of it.

Notice that by taking the trace of the curvature of the
connection $\nabla\oplus\hat{\nabla}$ on $\P$ we can
express $\omega$ in terms of topological quantum numbers:
$$\omega={2\pi i}\frac{\left(
\frac{1}{6}\iota(\tilde{\theta})^3\mu_4(\P)+
\frac{1}{2}\iota(\tilde{\theta})^2\mu_3(\P)+
\iota(\tilde{\theta})\mu_2(\P)+
\mu_1(\P)\right)|_{\Lt}}{\frac{1}{24}\iota(\tilde{\theta})^4\mu_4(\P)+
\frac{1}{6}\iota(\tilde{\theta})^3\mu_3(\P)+
\frac{1}{2}\iota(\tilde{\theta})^2\mu_2(\P)+
\iota(\tilde{\theta}) \mu_1(\P) + \mu_0(\P)} .$$
Here we use the notation $\tilde{\theta}$ for
$\theta\oplus(-\hat{\theta})$,
 $\mu_k(\P)$ for $2k$-dimensional component of $\mu(\P)$.

In general, the Nahm transform defined above is not
bijective (even
in
commutative case, i.e. when $\theta=\hat{\theta}=0$).
However the commutative Nahm transform
is bijective if $\P$ is ``Poincare module'' (the module
corresponding
to the Poincare line bundle). Strictly speaking, the term
``Nahm
transform''
is used only in this situation. It is natural to define
the Nahm
transform
in noncommutative case using  an $\Adt$
module having the same topological numbers as Poincare
module. (We
will
prove in Section \ref{sec5} that the deformed Poincare
module can be
equipped
with constant curvature connection and hence it can be
used to define the Nahm transform.)
One should expect that in noncommutative
case the Nahm transform is a bijection (and its square is
$(-1)^*$).
We can give only heuristic proof of this conjecture.

{\bf Remark.}  The relation between topological quantum
numbers of
$\hat{R}$
and $R$ in the case when $\P$ is deformed Poincare module
is the same
as
in commutative case (the relation between discrete
quantities cannot
change under continuous deformation):
\begin{equation}
\begin{array}{l}
\mu(R)=p+\frac{1}{2}a_{ij}\, \alpha^{i}\wedge\alpha^{j}+q\,\,
\alpha^1\wedge
\alpha^2\wedge\alpha^3\wedge\alpha^4,\\[12pt]
\mu(\hat{R})=q-\frac{1}{4}
\epsilon^{ijkl}a_{ij}\, \beta_k\wedge\beta_l+
p\,\, \beta_1\wedge\beta_2\wedge\beta_3\wedge\beta_4
\end{array}
\end{equation}
where $\epsilon^{ijkl}$ is completely antisymmetric
tensor  and the bases $\alpha^i$ and
$\beta_i$ are the standard dual bases of $H^2(T,\Z)$ and
$H^2(\hat{T},\Z)$.

We now present the formulae which  relate  $\omega$,
$\hat{\omega}$,
$\theta$, $\hat{\theta}$, etc. Recall that in the
commutative case
$L_{\theta} \approx L_{\hat\theta}^{\prime}$. The
integral class
$\mu(\P) = \exp {\sum_k \alpha^{k} \wedge
\beta_k}$
 is the same as in the commutative case and it allows to
identify
$L_{\theta}^{\prime} \approx L_{\hat\theta}$ in the
noncommutative
case too.
It is convenient to think
 of the forms $\omega, \hat\omega, \sigma$ as of the
operators:
$$
\omega: L_{\theta} \to L_{\theta}^{\prime} \approx
L_{\hat\theta}, \,
\, \hat\omega: L_{\hat\theta} \to L_{\theta}, \,\, \sigma:
L_{\hat\theta} \to L_{\hat\theta}.
$$
In the same way the bivectors $\theta,
\hat\theta$
are viewed as skew-symmetric operators: $\theta:
L_{\hat\theta} \to
L_{\theta}$, $\hat\theta:
L_{\theta} \to L_{\hat\theta}$.
{}From  (\ref{eqrelchmu}) using the Wick
theorem we get:
\begin{equation}
\label{eqrelthom}
\begin{array}{l}
\mbox{\rm dim}\,\P= ~~
\sqrt{{\rm Det}(1 - \hat\theta\theta)},\\[12pt]
\omega = ~~\hat\theta ( 1 - \theta\hat\theta)^{-1} \\[12pt]
\hat{\omega} = ~~- \theta ( 1 - \hat\theta\theta)^{-1}\\[12pt]
\sigma = ~~( 1 - \hat\theta\theta)^{-1}\\[12pt]
\end{array}
\end{equation}

\section{Definitions.}
\label{sec3}

We assume that all modules are Hermitian modules, all
connections
are Hermitian connections, and noncommutative tori $\At$
and $\Aht$
are
four dimensional. Let $(\P,\n,\hat{\n})$
be a finitely generated projective $\Adt$
module equipped with constant curvature connection
$\n\oplus\hat{\n}$.
The curvature of   $\n\oplus\hat{\n}$ is an element of
$\wedge^2(\Lt)'\oplus(\Lt\otimes\Lht)'\oplus\wedge^2(\Lht)'$.
We denote by $\omega_{\alpha\beta}$ the $\wedge^2(\Lt)'$
part of the
curvature of $\n\oplus\hat{\n}$, by $\sigma_{\alpha\mu}$ the
$(\Lt\otimes\Lht)'$ part, and by $\hat{\omega}_{\mu\nu}$ the
$\wedge^2(\Lht)'$
part.
We fix an inner product on $\Lt$. The inner
product on $\Lht$ is obtained from the inner product on
$\Lt$ via
the pairing $\sigma_{\alpha\mu}$. Let $\{\alpha_i\}$ be an
orthonormal basis
of $\Lt$.

\begin{defi}
\label{def1}
A connection $\CR$ on  $\At$-module $R$ is called
$\P$-irreducible
iff
there exists an inverse operator $G$ to the Laplacian
$\Delta=\sum_{i}(\CR_{\alpha_i}+\n_{\alpha_i})(\CR_{\alpha_i}+\n_{\alpha_i})$
and $G$ is bounded operator.
\end{defi}

\begin{lem}
\label{lem1}
If $\CR$ is $\P$-irreducible connection on $R$ and $F^+
+\omega^+\cdot 1=0$
then $\mbox{\rm ker}(D^+)=0$, and
\begin{equation}
\label{eq11}
D^- D^+=\Delta.
\end{equation}
\end{lem}
{\bf Proof:} The proof is the same as in the commutative case.
$\Box$

Now we can define a noncommutative Nahm transform. Let
$R$ be a
projective
Hermitian module over $\At$ with $\P$-irreducible
connection $\CR$
such that
its curvature $F$ satisfies the condition $F^+
+\omega^+\cdot 1=0$.
Denote by $\hat{R}$ the closure
of the kernel of
$D^-: (R\otimes_{{\cal A}_{\theta}}{\cal P})\otimes
S^-\rightarrow
(R\otimes_{{\cal A}_{\theta}}{\cal P})\otimes S^+.$
Notice that $(R\otimes_{{\cal A}_{\theta}}{\cal
P})\otimes S^-$ is
Hermitian $\Aht$ module and that $D^-$ commutes with
$\Aht$ action.
Therefore, $\hat{R}$ is a Hermitian $\Aht$ module
(submodule of
$(R\otimes_{{\cal A}_{\theta}}{\cal P})\otimes S^-$).
Let us denote by $P$ the
projection operator (with respect to the $\Aht$ Hermitian
inner
product)
from $(R\otimes_{{\cal A}_{\theta}}{\cal P})\otimes S^-$
onto $\hat{R}$.
In other words, $P$ is Hermitian, $P^2=P$, and $\mbox{\rm
Im}\,P=\hat{R}$.
Its existence is proven in the theorem \ref{thm1} below.
We denote by $\ChR$ the composition $P\circ \hat{\n}$.

\begin{thm}
\label{thm1}
$\hat{R}$ is a finitely generated projective Hermitian
$\Aht$ module
and
$\ChR$ is a Hermitian $\Aht$ connection on $\hat{R}$.
\end{thm}
{\bf Proof:}
The projection operator on the kernel of $D^-$ can be defined
by the following explicit formula $P=1-D^+ G D^-$.
We can check that $P$ is hermitian, $P^2=P$, and
$\mbox{\rm Im}\,P=\mbox{\rm Ker}\,D^-$ by means of formal
algebraic
manipulations using $D^-D^+=\Delta$.
We claim that $P$ is ``compact'' operator over $\Aht$
(that is a
limit
of the operators of the type $\sum_i x_i(y_i,\cdot)$).
This follows immediately as usual from the fact that $D^+$
admits a parametrix (over $\Aht$, see Appendix A)
and $G D^-$ is left inverse to $D^+$.
Since $P$ is ``compact'' over $\Aht$ it follows from the
general
theory (see
Appendix A or \cite{Bl}) that $P=\sum_j u_j(v_j,\cdot)$ is
a projection on a finitely generated projective $\Aht$
module which
proves the first statement. $\hat{R}$ inherits $\Aht$
valued inner
product
from $(R\otimes_{{\cal A}_{\theta}}{\cal P})\otimes S^-$.

The second statement follows immediately from the fact
that $P$
commutes
with the action of $\Aht$ (since $D^+$, $D^-$, and $G$
commute with
$\Aht$ action). $\Box$

\begin{defi}
\label{def3}
The noncommutative Nahm transform $\N$ of $(R,\CR)$ is a pair
$(\hat{R},\ChR)$
  of projective Hermitian $\Aht$ module $\hat{R}$
and connection $\ChR$.
\end{defi}

\section{Properties of noncommutative Nahm transform.}
\label{sec4}

Now we can find a formula for the curvature of $\ChR$.

\begin{prop}
\label{thm2}
We have the following formula for the curvature $\hat{F}$ of
connection $\ChR$
\begin{equation}
\label{eq15}
\hat{F}_{\mu\nu}=\hat{\omega}_{\mu\nu}+PG\left(\sum_{\alpha,\beta}
(\Gamma^\alpha\Gamma^\beta-\Gamma^\beta\Gamma^\alpha)\sigma_{\alpha\mu}
\sigma_{\beta\nu}\right).
\end{equation}
\end{prop}

\begin{rem}
\label{rem1}
It follows from (\ref{eq11}) that $D^-D^+$ commutes with
$\Gamma^\alpha$.
Therefore $\Gamma^\alpha$ commutes with G. We will use
this in the
proof.
\end{rem}

{\bf Proof:} Since the Lie algebra $\Lht$ is commutative the
curvature
$\hat{F}_{\mu\nu}=[\ChR_\mu,\ChR_\nu]$.
The definition of $\ChR$ gives us that
$\hat{F}_{\mu\nu}=P\hat{\n}_\mu P \hat{\n}_\nu-
P \hat{\n}_\nu P\hat{\n}_\mu$. Let us simplify this expression
using the definition of $P=1-D^+ G D^-$. If we replace
the middle $P$
by this
expression we obtain
\begin{equation}
\label{eq16}
\begin{array}{l}
\hat{F}_{\mu\nu}=P [\hat{\n}_\mu,\hat{\n}_\nu]+P
(\hat{\n}_\nu D^+ G D^-\hat{\n}_\mu-\hat{\n}_\mu D^+ G
D^-\hat{\n}_\nu)
=\\[12pt]
{}~~~~~~~~~~=\hat{\omega}_{\mu\nu}+P
(\hat{\n}_\nu D^+ G D^-\hat{\n}_\mu-\hat{\n}_\mu D^+ G
D^-\hat{\n}_\nu).
\end{array}
\end{equation}
Next, let us notice that $PD^+=0$ and $D^-$ equals zero
on $\hat{R}$.
Therefore, we can rewrite (\ref{eq16}) as
\begin{equation}
\label{eq17}
\hat{F}_{\mu\nu}=\hat{\omega}_{\mu\nu}+P
([\hat{\n}_\nu,D^+]G[D^-,\hat{\n}_\mu]-[\hat{\n}_\mu,D^+]G[D^-,\hat{\n}_\nu]).
\end{equation}
To proceed further we need the following
commutation relation
\begin{equation}
\label{eq18}
[D,\hat{\n}_\lambda]=\sum_{\alpha}\Gamma^{\alpha}\sigma_{\alpha\lambda}.
\end{equation}
(The proof is straightforward calculation using the definition
of
$D= \hfill \\
\sum_{\alpha}\Gamma^{\alpha}(\CR_\alpha+\n_{\alpha})$ and
the fact
that
$\CR$ commutes with $\hat{\n}$.)

The immediate consequence of (\ref{eq18}) is that
$$
\hat{F}_{\mu\nu}=\hat{\omega}_{\mu\nu}+P
\left(-\{\sum_{\beta}\Gamma^{\beta}\sigma_{\beta,\nu}\}G
\{\sum_{\alpha}\Gamma^{\alpha}\sigma_{\alpha\mu}\}+
\{\sum_{\alpha}\Gamma^{\alpha}\sigma_{\alpha\mu}\}G
\{\sum_{\beta}\Gamma^{\beta}\sigma_{\beta\nu}\}\right).
$$
Now the theorem follows from the remark that $\Gamma^\lambda$
commutes with $G$ and very simple formal manipulations.
$\Box$\\

Next, we would like to compute $\hat{F}^+$
(the self-dual part of the curvature).
Let us remind that the inner product on $\Lht$ came from a
non-degenerate
pairing between $\Lt$ and $\Lht$ given by
$\sigma_{\alpha\mu}$.

\begin{lem}
\label{lem3}
The selfdual part of $\left(\sum_{\alpha,\beta}
(\Gamma^\alpha\Gamma^\beta-\Gamma^\beta\Gamma^\alpha)\sigma_{\alpha\mu}
\sigma_{\beta\nu}\right)$ on $S^-$ is equal to zero.
\end{lem}
{\bf Proof:} Let us choose an orthonormal basis
$\{\alpha_i\}$ in
$\Lt$.
Let $\{\mu_j\}$ be the
dual basis in $\Lht$ with respect to the pairing given by
$\sigma_{\alpha\mu}$. Then, the element
$$\left(\sum_{i,j}
(\Gamma^{\alpha_i}\Gamma^{\alpha_j}-\Gamma^{\alpha_j}\Gamma^{\alpha_i})
\sigma_{\alpha_i,\mu_k}\sigma_{\alpha_j,\mu_l}\right)=
\Gamma^{\alpha_k}\Gamma^{\alpha_l}-\Gamma^{\alpha_l}\Gamma^{\alpha_k}
$$
operator
It is well-known that this element is antiselfdual on $S^-$.
Thus, the selfdual part of it
is equal to zero. $\Box$\\

As an immediate corollary we obtain that noncommutative Nahm
transform
is similar to a commutative Nahm transform in the
following relation

\begin{thm}
\label{thm2.5}
Let $(R,\CR)$ be a
projective Hermitian module over $\At$ with $\P$-irreducible
connection $\CR$ which satisfy the condition
$F^++\omega^+ \cdot
1=0$,
where $F$ is the curvature of $\CR$. Let $(\hat{R},\ChR)$ be a
noncommutative
Nahm transform of $(R,\CR)$. Then the curvature $\hat{F}$
of $\ChR$
satisfies
the equation $\hat{F}^+-\hat{\omega}^+\cdot 1=0$.
\end{thm}
{\bf Proof:} The statement immediately follows from
Proposition
\ref{thm2} and
the previous lemma. $\Box$\\

Let ${\hat{R}}^*$ be the left $\Aht$ module dual to
$\hat{R}$, {\it
i.e.},
 ${\hat{R}}^*=\mbox{\rm Hom}_{\Aht}(\hat{R},\Aht)$.
Notice that as a
vector
space ${\hat{R}}^*$ is isomorphic to $\hat{R}$
since $\hat{R}$ is a projective
Hermitian $\Aht$ module. Consider the tensor product
$\hat{R}\otimes_{\Aht}{\hat{R}^*}$.
Since $\hat{R}$ is a finitely generated projective
$\Aht$ module
the algebra $\hat{R}\otimes_{\Aht}{\hat{R}^*}$
is naturally isomorphic to the algebra
$\mbox{\rm End}_{\Aht}(\hat{R})$. Let $e$ be an identity
element in
$\mbox{\rm End}_{\Aht}(\hat{R})$. By abuse of notation we
denote its
image in $\hat{R}\otimes_{\Aht}{\hat{R}^*}$ by the same
letter $e$.
\begin{rem}
\label{rem2}
Notice that $e$ is a finite sum $\sum_i x_i\otimes y_i$, where
$x_i\in{\hat{R}}$ and $y_i\in{\hat{R}^*}$, because
$\hat{R}$ is a
finitely
generated projective Hermitian module over $\Aht$.
\end{rem}
The module $\hat{R}$ was defined as a submodule of
$(\RtP)\otimes S^-$. Therefore, we have a canonical embedding
$\hat{R}\otimes_{\Aht}{\hat{R}^*}$ into
 $\left((\RtP)\otimes S^-\right)\otimes_{\Aht}{\hat{R}^*}$.
Denote by $\Psi$ the image of $e$ in
$\left((\RtP)\otimes S^-\right)\otimes_{\Aht}{\hat{R}^*}$.
Let us notice that the module
$\left((\RtP)\otimes S^-\right)\otimes_{\Aht}{\hat{R}^*}$ is
canonically
isomorphic to $(\RtP\otimes_{\Aht}{\hat{R}^*})\otimes
S^-$. We use
the latter one everywhere instead of the former.

Denote by $R^*$ the left $\Aht$ module dual to ${R}$,
$R^*=\mbox{\rm Hom}_{\At}({R},\At)$. Any element $f\in
R^*$ gives us
a map
from $(\RtP\otimes_{\Aht}{\hat{R}^*})\otimes S^-$ to
$(\P\otimes_{\Aht}{\hat{R}^*})\otimes S^-$.
$$
(\RtP\otimes_{\Aht}{\hat{R}^*})\otimes S^-\ni
(x\otimes p \otimes y)\otimes s\mapsto (f(x)p\otimes
y)\otimes s
\in (\P\otimes_{\Aht}{\hat{R}^*})\otimes S^-,
$$
where $x\in R$, $p\in\P$, $y\in{\hat{R}^*}$, and $s\in S^-$.
We denote this map by $\bf f$.

Notice, that $G$ (inverse to $D^-D^+$) commutes with the
action of
$\Aht$.
Therefore, it acts on
$(R\otimes_{\At}\P\otimes_{\Aht}{\hat{R}^*})\otimes S^-$.
We would like to consider a canonical element $G\Psi\in
(R\otimes_{\At}\P\otimes_{\Aht}{\hat{R}^*})\otimes S^-$.
Strictly speaking the spinor spaces in the definitions of
$D_{R}$ and
$D_{\hat{R}^*}$ are different (one of them is constructed
using
$\Lt$,
another one using  $\Lht$). However, we may use
$\sigma_{\alpha\mu}$ to
identify Euclidean spaces $\Lt$ and $\Lht$ and hence the
corresponding spinor
spaces. Thus we can consider the Dirac operator
$D^-_{\hat{R}^*}$ as
a map
from $(\P\otimes_{\Aht}{\hat{R}^*})\otimes S^-$ to
$(\P\otimes_{\Aht}{\hat{R}^*})\otimes S^+$. Notice that this
identification does not respect the duality used in
(\ref{eqrelthom}).

\begin{prop}
\label{prop2}
For any element $f\in R^*$ the element ${\bf f}(G\Psi)\in
(\P\otimes_{\Aht}{\hat{R}^*})\otimes S^-$ lies in the
kernel of
$D^-_{\hat{R}^*}$.
\end{prop}
{\bf Proof:}
First, we need two lemmas.
\begin{lem}
\label{lem4}
Let $\{\alpha_i\}$ be an orthonormal basis in $\Lt$ and
$\{\mu_i\}$
be the
dual basis (also orthonormal) in $\Lht$ (the pairing
between $\Lt$
and $\Lht$
is given by $\sigma_{\alpha\mu}$). Then we have
\begin{equation}
\label{eq21}
[\hat{\n}_{\mu_i},G]=2G(\CR_{\alpha_i}+\n_{\alpha_i})G.
\end{equation}
\end{lem}
{\bf Proof:} Recall that $G$ was defined as an inverse
operator to
$D^-D^+$.
 From Lemma \ref{lem1} we know that
$D^-D^+=\sum_i(\CR_{\alpha_i}+\n_{\alpha_i})^*(\CR_{\alpha_i}+\n_{\alpha_i})=
\sum_i(\CR_{\alpha_i}+\n_{\alpha_i})(\CR_{\alpha_i}+\n_{\alpha_i})$,
since
all connections are by definition Hermitian, {\it i.e.},
selfadjoint.
 From (\ref{eq18}) we obtain
\begin{equation}
\label{eq22}
[D^-D^+,\hat{\n}_{\mu_j}]=[\sum_i(\CR_{\alpha_i}+\n_{\alpha_i})
(\CR_{\alpha_i}+\n_{\alpha_i}),\hat{\n}_{\mu_j}]=
2(\CR_{\alpha_j}+\n_{\alpha_j}),
\end{equation}
since $\sigma_{\alpha_k,\mu_l}=\delta_{k,l}$. Multiplying
the formula
(\ref{eq22}) by $G$ from the left and by $G$ from the
right and using
the fact
that $G$ is inverse to $D^-D^+$, we obtain (\ref{eq21}).
$\Box$\\

Let $e=\sum_i x_i\otimes y_i$ be an element of
$\hat{R}\otimes_{\Aht}{\hat{R}}^*$ as in Remark
\ref{rem2}. Then, any
element
$z\in\hat{R}^*$ can be written as
$$z=\sum_i (z,x_i)y_i$$
where $(z,x_i)=z(x_i)$ is an element of $\Aht$ and we
think about $z$
as
a homomorphism from $\hat{R}$ to $\Aht$. The goal of the
next lemma
is to
describe connection $(\ChR)^*$ on $\hat{R}^*$.
\begin{lem}
\label{lem5}
For any smooth element $z\in{\hat{R}^*}$ we have
\begin{equation}
\label{eq23}
(\ChR_{\mu_i})^*z=-\sum_{k,l}(z,\hat{\n}_{\mu_i}x_k-
\Gamma^{\alpha_l}\Gamma^{\alpha_i}(\CR_{\alpha_l}+\n_{\alpha_l})Gx_k)y_k,
\end{equation}
where the basises $\{\alpha_i\}$ and $\{\mu_i\}$ are
chosen as in
Lemma
\ref{lem4}.
\end{lem}
{\bf Proof:} The proof is the following  tedious trivial
calculation.
$$
\begin{array}{l}
(\ChR_{\mu_i})^*z=\sum_k ((\ChR_{\mu_i})^*z,x_k)y_k=
-\sum_k (z,\ChR_{\mu_i}x_k)y_k
=-\sum_k (z,P\hat{\n}_{\mu_i}x_k)y_k=\\[12pt]
-\sum_k (z,(1-D^+GD^-)\hat{\n}_{\mu_i}x_k)y_k=
-\sum_k
(z,\hat{\n}_{\mu_i}x_k-D^+G[D^-,\hat{\n}_{\mu_i}]x_k)y_k,
\end{array}
$$
since $D^-x_k=0$. From (\ref{eq18}) and the choice of the
basises
$\{\alpha_i\}$ and $\{\mu_i\}$ it follows that
$[D^-,\hat{\n}_{\mu_i}]=\Gamma^{\alpha_i}$. Therefore, we
obtain
$$
\begin{array}{c}
(\ChR_{\mu_i})^*z=
-\sum_k
(z,\hat{\n}_{\mu_i}x_k-D^+G\Gamma^{\alpha_i}x_k)y_k=\\[12pt]
{}~~~~=-\sum_{k,l}
(z,\hat{\n}_{\mu_i}x_k-\Gamma^{\alpha_l}
\Gamma^{\alpha_i}(\CR_{\alpha_l}+\n_{\alpha_i})Gx_k)y_k,
\end{array}
$$
since $G$ commutes with $\Gamma^\alpha$ (and we replaced
$D^+$ by its
definition $D^+=\sum_l
\Gamma^{\alpha_l}(\CR_{\alpha_l}+\n_{\alpha_l})$).
$\Box$\\

Now we  prove the proposition.
Let us choose the bases $\{\alpha_i\}$ and $\{\mu_i\}$ as
in lemma
\ref{lem4}.
First, recall that
$D^-_{\hat{R}^*}=\sum_j \Gamma^{\mu_j}\left(\hat{\n}_{\mu_j}+
(\ChR_{\mu_j})^*\right)$. Since we identified the spinors
for $\Aht$
with
the spinors for $\At$ $\Gamma^{\mu_j}=\Gamma^{\alpha_j}$ and
$D^-_{\hat{R}^*}=\sum_j
\Gamma^{\alpha_j}\left(\hat{\n}_{\mu_j}+
(\ChR_{\mu_j})^*\right)$. Second, recall that
$\Psi=\sum_k x_k\otimes
y_k$,
therefore
$${\bf f}(G\Psi)={\bf f}(G\sum_k x_k\otimes y_k)=
\sum_k {\bf f}(Gx_k\otimes y_k),$$
since $G$ acts only on the first argument. Third, notice that
$D^-_{\hat{R}^*}$ commutes with ${\bf f}$ therefore we obtain
$$
D^-_{\hat{R}^*}({\bf f}(G\Psi))={\bf f}\left[\sum_{j,k}
\Gamma^{\alpha_j}\left(\hat{\n}_{\mu_j}+
(\ChR_{\mu_j})^*\right)(Gx_k\otimes y_k)\right].
$$
We continue our manipulations
\begin{equation}
\label{eq24}
D^-_{\hat{R}^*}({\bf f}(G\Psi))={\bf f}\left[\sum_{j,k}
\Gamma^{\alpha_j}\left(\hat{\n}_{\mu_j}Gx_k\otimes y_k+
Gx_k\otimes(\ChR_{\mu_j})^* y_k\right)\right].
\end{equation}
Using lemma \ref{lem5} we can rewrite
$$
\begin{array}{l}
\sum_k Gx_k\otimes(\ChR_{\mu_j})^* y_k=
\\[12pt]
=\sum_k Gx_k\otimes(
-\sum_{l,m}(y_k,\hat{\n}_{\mu_j}x_m-
\Gamma^{\alpha_l}\Gamma^{\alpha_j}(\CR_{\alpha_l}+\n_{\alpha_l})Gx_m)y_m)=
\\[12pt]
=-\sum_{k,l,m} Gx_k ((y_k,\hat{\n}_{\mu_j}x_m-
\Gamma^{\alpha_l}\Gamma^{\alpha_j}(\CR_{\alpha_l}+\n_{\alpha_l})Gx_m)
\otimes y_m=\\[12pt]
=-\sum_{l,m}G(\hat{\n}_{\mu_j}x_m-
\Gamma^{\alpha_l}\Gamma^{\alpha_j}(\CR_{\alpha_l}+\n_{\alpha_l})Gx_m)\otimes
y_m=\\[12pt]
=-\sum_{k,l} \left(G\hat{\n}_{\mu_j}x_k-
\Gamma^{\alpha_l}\Gamma^{\alpha_j}G(\CR_{\alpha_l}+\n_{\alpha_l})Gx_k\right)
\otimes y_k,
\end{array}
$$
in the last line we replaced everywhere $m$ by $k$.
Substituting
this
into the  formula (\ref{eq24}) we get
\begin{equation}
\label{eq25}
\begin{array}{l}
D^-_{\hat{R}^*}({\bf f}(G\Psi))={\bf f}\\[12pt]
\left[\sum_{j,k}
\Gamma^{\alpha_j}\left(\hat{\n}_{\mu_j}Gx_k\otimes y_k
-\sum_{k,l} \left(G\hat{\n}_{\mu_j}x_k-
\Gamma^{\alpha_l}\Gamma^{\alpha_j}G(\CR_{\alpha_l}+\n_{\alpha_l})Gx_k\right)
\otimes y_k\right)\right]\\[12pt]
={\bf
f}\left[\sum_k\left(\sum_j\Gamma^{\alpha_j}[\hat{\n}_{\mu_j},G]+
\sum_{j,l}\Gamma^{\alpha_j}\Gamma^{\alpha_l}\Gamma^{\alpha_j}
G(\CR_{\alpha_l}+\n_{\alpha_l})G\right)x_k\otimes
y_k\right].
\end{array}
\end{equation}
Using lemma \ref{lem4} we substitute
$2G(\CR_{\alpha_j}+\n_{\alpha_j})G$ instead of
$[\hat{\n}_{\mu_j},G]$
and obtain
\begin{equation}
\label{eq26}
D^-_{\hat{R}^*}({\bf f}(G\Psi))=
{\bf f}\left[\sum_{k,l}\left(2\Gamma^{\alpha_l}+\sum_j
\Gamma^{\alpha_j}\Gamma^{\alpha_l}\Gamma^{\alpha_j}\right)
G(\CR_{\alpha_l}+\n_{\alpha_l})Gx_k\otimes y_k\right].
\end{equation}
The proposition follows from the fact that
$2\Gamma^{\alpha_l}+\sum_j
\Gamma^{\alpha_j}\Gamma^{\alpha_l}\Gamma^{\alpha_j}$
equals zero for all $l$, since the dimension is four. $\Box$

We interpret this proposition as existence of $\At$
homomorphism
from $R^*$ to the kernel of $D^-_{\hat{R}^*}$. If we knew that
${\hat{R}^*}$ is $\P$-irreducible we could say that we
have a $\At$
homomorphism from $R^*$ to $\widehat{{\hat{R}^*}}$.

\section{Examples.}
\label{sec5}

In our previous discussion we assumed that we have a
module $\cal P$
over
${\A}_{\theta\oplus(-\hat{\theta})}$ with some
properties. In this
section we
give examples of such modules. Although the modules are
quite simple
the
constructions of connections with desired properties are quite
technical,
and tedious.

At first we consider so called elementary
finitely generated projective
modules over
${\A}_{\theta\oplus{(-\hat{\theta})}}$ (see \cite{Rf1}).
The second example shows how to deform the commutative
Poincare
module.
Essentially we will show that the commutative Poincare module
can be viewed as an elementary finitely generated
projective module
(in the language of \cite{Rf1}).
This allows us to deform Poincare module and constant
curvature connection on it using the constructions of the
first part of the section.

In the first part we follow mainly the exposition of paper
\cite{Rf1} with
some minor modifications.
We can think about
$\theta$ and $\hat{\theta}$ as skew-symmetric bilinear
forms on
$\Z^4$. We  embed $\Z^4$ in a usual way into $\R^4$
and extend
the forms $\theta$ and $\hat{\theta}$ to be
skew-symmetric bilinear
forms on $\R^4$.
For simplicity we  assume that $\theta$ and
$\hat{\theta}$ are
non-degenerate symplectic forms.
We can take their direct sum $\theta\oplus(-\hat{\theta})$
and consider it as  a two-form on $\R^8$. Now take
another $\R^8=\R^4\oplus(\R^4)'$ with a canonical
skew-symmetric
bilinear form $\Omega$ given by $\Omega
\left(\left(x_1,y_1\right),\left(x_2,y_2\right)\right)=
\langle y_2,x_1\rangle-\langle y_1,x_2\rangle$, where
$x_1,x_2\in\R^4$ and
$y_1,y_2\in(\R^4)'$. Let $\eta$ be an arbitrary integral
two form
(by definition, it takes integer values on the lattice
$\Z^8$) on
$\R^8$.
Let $T$ be a linear map from $\R^8$ to
$\R^4\oplus(\R^4)^{\prime}$
such that $\theta\oplus(-\hat{\theta})+\eta=T^*(\Omega)$.
Such map always exists but
it is never unique
(any two maps of this kind
are conjugate by an element from a symplectic group).
Now we can describe some examples of modules $\P$.

We can realize $\P$ as a space of functions on $\R^4$;
the smooth part $\P^{smooth}=S(\R^4)$ is the space of
Schwartz functions on $\R^4$. Let us describe the left action
of $\At$ and the right action of $\Aht$ (which commute
with each
other).
First, notice that $\R^4\oplus(\R^4)'$ acts on functions as
follows:
$\left(\left(x,y\right)f\right)\left(z\right)=e^{2\pi i
\langle y,z\rangle}f(z+x)$.
The left
action of an element $U_\nu\in \At$, $\nu\in\Z^4$ is given by
$$(U_\nu f)(z)=(T(i_1(\nu))f)(z),$$
where $i_1$ is the canonical inclusion of
$\Z^4\hookrightarrow \R^4
\hookrightarrow \R^4\oplus\R^4=\R^8$ in the first $\R^4$,
{\it i.e.},
$i_1(\nu)=(\nu,0)$.
Similarly, we define $i_2$ as the canonical inclusion of
$\Z^4$ in
the second
$\R^4$ in $\R^8$, {\it i.e.}, $i_2(\nu)=(0,\nu)$.
We define the right action of $\hat{U}_\nu\in \Aht$,
$\nu\in\Z^4$ as
$$(\hat{U}_\nu f)(z)=(T(i_2(\nu))f)(z).$$
A straightforward calculation shows that in this way we
actually get
an $(\At,\Aht)$ module. For the proof that $\P$ is projective
$\Adt$ module see \cite{Rf1}.

Next we define connections $\n$ and $\hat{\n}$ on $\P$.
Notice that
we can identify the Lie algebra $\Lt$ with $(\R^4)'$. If
$\alpha\in (\R^4)'$ then the corresponding derivation
$\delta_\alpha$
acts
on $U_\nu$, $\nu\in\Z^4$ as a multiplication by
$2\pi i\langle \alpha,\nu\rangle$.
Similarly, we can identify the Lie algebra $\Lht$ with
$(\R^4)'$.
For $\alpha\in (\R^4)'$, we have
$\delta_\alpha(\hat{U}_\nu)=2\pi i
\langle \alpha,\nu\rangle \hat{U}_\nu$.

Now, let us define for any $(x,y)\in \R^4\oplus(\R^4)'$
an operator
$Q_{(x,y)}$ on the smooth functions on $\R^4$ as follows
$$(Q_{(x,y)}f)(z)=2\pi i \langle y,z\rangle f(z)+
\frac{d(f(z+tx))}{dt}|_{t=0}.$$
The straightforward calculation  shows that
$$\begin{array}{l}
[Q_{(x,y)},U_\nu]=2\pi i
\Omega((x,y),T((\nu,0)))U_\nu,\\[12pt]
[Q_{(x,y)},\hat{U}_\nu]=2\pi i
\Omega((x,y),T((0,\nu)))\hat{U}_\nu.
\end{array}$$

We define connection $\n_\alpha$ as $Q_{V_1(\alpha)}$, where
$V_1$ is the unique map from $\Lt=(\R^4)'$ such that
$$\Omega(V_1(\alpha),T((\nu,\mu)))=\langle
\alpha,\nu\rangle,$$
where $\alpha\in \Lt=(\R^4)'$, and $\nu,\mu\in\R^4$.
Since $\Omega$ is a non-degenerate bilinear form, $T$ is an
isomorphism and the existence and uniqueness of $V_1$ is
obvious.
{}From this definition it is clear that $\n$ commutes
with $\Aht$.
Similarly, we define $\hat{\n}_\alpha$ as
$Q_{V_2(\alpha)}$, where
$V_2$ is the unique map from $\Lht=(\R^4)'$ such that
$$\Omega(V_2(\alpha),T((\nu,\mu)))=\langle
\alpha,\mu\rangle,$$
where $\alpha\in \Lht=(\R^4)'$, and $\nu,\mu\in\R^4$.
{}From this definition we immediately see that $\hat{\n}$
commutes
with
$\At$.

To compute the curvature of the connections $\n$ and
$\hat{\n}$ and
to compute
 the commutators $[\n_\alpha,\hat{\n}_\beta]$ we need to
know the
commutator
$[Q_{(x_1,y_1)},Q_{(x_2,y_2)}]$.
It is easy to see that
$$[Q_{(x_1,y_1)},Q_{(x_2,y_2)}]=2\pi i
\Omega\left((x_1,y_1),(x_2,y_2)\right)\Id.$$
As an immediate corollary we obtain that connections $\n$ and
$\hat{\n}$ have
constant curvature and that $[\n_\alpha,\hat{\n}_\beta]=
2\pi i \Omega(V_1(\alpha),V_2(\beta))$.  It is not hard
to check that
generically this pairing will be non-degenerate.
Therefore, for
generic
$\theta$ and $\hat{\theta}$
we get many non-trivial examples of module $\P$.

{}The most interesting example of module $\P$ can be
obtained by deforming  the ``Poincare module''  -
the space of
sections
of Poincare line bundle (see \cite{bbr}, \cite{Dk}).

 {}Let us remind one of the possible definitions of the
Poincare
module
Let $\Z^4$ be a lattice
in $\R^4$. Let $(\Z^4)'$ be the dual lattice in $(\R^4)'$. The
Poincare
module $\P$ consists of
functions $f(x,\hat{x})$ on $\R^4\oplus(\R^4)'$
which satisfy the following condition:
\begin{equation}
\label{Poicbun}
f(x+\lambda,\hat{x}+\hat{\lambda})=e^{-2\pi
i\langle\hat{\lambda},x\rangle}
f(x,\hat{x}),
\end{equation}
for all $\lambda\in\Z^4$ and $\hat{\lambda}\in(\Z^4)'$.
The algebra of functions on the torus considered as the
algebra of
periodic functions on  $\R^4\oplus(\R^4)'$; it acts on
the module by
 multiplication.

We will use another construction of Poincare module (this
construction is similar to the construction of modules over
two-dimensional torus given in \cite{Ho}).

Notice that for fixed $\hat{x}$  the function $f(x,\hat{x})$
is a periodic function in $x$ which can be written
as a Fourier series with $\hat{x}$-dependent coefficients:
\begin{equation}
\label{Four}
f(x,\hat{x})=\sum_{\mu\in (\Z^4)'} e^{2\pi
i\langle\mu,x\rangle}
c_{\mu}(\hat{x}).
\end{equation}
The coefficients $c_{\mu}(\hat{x})$ are smooth functions
of $\hat{x}$ if the original function was smooth since
they are given
by the integrals over torus (which is compact).
For convenience we denote $c_0(\hat{x})$ by $\phi(\hat{x})$.
The property (\ref{Poicbun}) of function $f(x,\hat{x})$
gives us that
$$c_{\mu}(\hat{x})=c_0(\hat{x}+\mu).$$
Therefore, we can rewrite (\ref{Four}) as
\begin{equation}
\label{decomp}
f(x,\hat{x})=\sum_{\mu\in (\Z^4)'} e^{2\pi
i\langle\mu,x\rangle}
\phi(\hat{x}+\mu).
\end{equation}
Moreover, it is not hard to see that function
$f(x,\hat{x})$ given by
the
formula (\ref{decomp}) is smooth if and only if the function
$\phi(\hat{x})$
belongs to the Schwartz space.

The algebra ${\cal A}_0$ is generated by the elements
$U_{\mu}=e^{2\pi i\langle\mu,x\rangle}$ and
$\hat{U}_{\nu}=e^{2\pi i\langle\nu,\hat{x}\rangle}$, where
$\nu\in\Z^4$ and $\mu\in (\Z^4)'$. We can rewrite the
action of the
operators $U_{\mu}$ and $\hat{U}_{\nu}$ in terms of their
action on
$\phi(\hat{x})$. One easily obtains
\begin{equation}
\label{actionn}
\hat{U}_{\nu}(\phi(\hat{x}))=e^{2\pi
i\langle\nu,\hat{x}\rangle}\phi(\hat{x})
{}~~~~~~\mbox{\rm and}~~~~~~~
U_{\mu}(\phi(\hat{x}))=\phi(\hat{x}-\mu).
\end{equation}

Thus, we realized the Poincare module as the space of
functions on
$(\R^4)'$
with the action given by the formula (\ref{actionn}).

The constant curvature connection on the Poincare line
bundle is
given by
$$\nabla=d+2\pi i\langle \hat{x},dx\rangle$$
in the first realization of Poincare module.
We can easily rewrite it in terms of its
action on the function $\phi(\hat{x})$. If $y\in\R^4$ and
$\hat{y}\in(\R^4)'$
 then
\begin{equation}
\label{connectn}
\nabla_y\phi(\hat{x})=2\pi i\langle
y,\hat{x}\rangle\phi(\hat{x})~~~~~~
\mbox{\rm and}~~~~~~~
\nabla_{\hat{y}}\phi(\hat{x})=\frac{d\,\phi(\hat{x}+t\hat{y})}{dt}|_{t=0}.
\end{equation}

One can also easily rewrite the Hermitian inner product
$\langle f(x,\hat{x}),\tilde{f}(x,\hat{x})\rangle$ in terms of
$\phi(\hat{x})$ and $\tilde{\phi}(\hat{x})$. In
particular, the trace of the inner product
of $f(x,\hat{x})$ and $\tilde{f}(x,\hat{x})$
equals
\begin{equation}
\label{innerpr}
\tau(\langle f(x,\hat{x}),\tilde{f}(x,\hat{x})\rangle)=
\int_{-\infty}^\infty
\phi(\hat{x})\overline{\tilde{\phi}(\hat{x})}d\hat{x}.
\end{equation}

{}From the above discussion we see that Poincare module
fits very
well in the
picture in the first example. More precisely, the map $T$
from the
first
example is
$$T(x,y)=(x,y).$$
Here, we identified the dual space $(\R^4)'$ with $\R^4$
using some
fixed standard
integral bases. We can easily calculate $\theta$,
$\hat{\theta}$, and
$\eta$ and we obtain that $\theta=\hat{\theta}=0$ and
$\eta$ is the two form on $\R^8$ given by
$$\eta((x_1,y_1),(x_2,y_2))=\langle y_2,x_1\rangle-
\langle y_1,x_2\rangle,$$
where
$x_i\in\R^4\stackrel{(Id,0)}{\hookrightarrow}\R^4\oplus\R^4=\R^8$
and
$y_i\in\R^4\stackrel{(0,Id)}{\hookrightarrow}\R^4\oplus\R^4=\R^8$,
$i=1,2$.
The inner product $\langle\cdot,\cdot\rangle$ comes from
the above
identification of $\R^4$ with $(\R^4)'$. It is obvious
that $\eta$ is
integral two form in this case.

We can easily deform the map $T$ so that the form
$\eta$ will be preserved but the forms $\theta$ and
$\hat{\theta}$
will
 be deformed. In this way we obtain a deformation of the
commutative
Poincare module. One can easily check that this
deformation is always
possible for small $\theta$ and $\hat{\theta}$. Using the
above
construction we obtain a constant curvature connection
in the deformed Poincare module.

Explicitly, the equation (\ref{actionn}) is deformed to:
\begin{equation}
\label{actionnn}
\hat{U}_{\nu}(\phi(\hat{x}))=e^{2\pi
i\langle C\nu,\hat{x}\rangle}\phi(\hat{x} - A  \nu)
{}~~~~~~\mbox{\rm and}~~~~~~~
U_{\mu}(\phi(\hat{x}))=e^{2\pi
i\langle B\mu,\hat{x}\rangle} \phi(\hat{x}-D \mu).
\end{equation}
where $A,B, C, D$ are the operators: $\R^4 \to (
\R^4)^{\prime}$
such that:
$$
B^{t} D - D^{t} B = \frac{\theta}{2\pi i}, \,\, C^{t} A -
A^{t} C =
\frac{\hat\theta}{2\pi i}, \,\, B^{t} A - D^{t} C\in {\rm
GL}(4, \Z)
$$

\section{Appendix A.}
\label{sec7}

In this section we will discuss under which conditions on
the module
$\P$
the kernel of $D^-$
(we assume that the kernel of $D^+$ is trivial)
is a finitely generated projective module over $\Aht$.
First, we will explain how to construct a parametrix to
$D^-$ ($D$).
More precisely, what we want to construct is a
``compact'' operator
$Q$ commuting with the action of $\Aht$ such that
$DQ=1+K_l$ and $QD=1+K_r$ and $K_l,K_r,DK_r,DK_l,K_rD,K_lD$
are ``compact'' operators. This is enough for all our
purposes.

We would like to construct a parametrix to $D:\,
E\otimes_{\At}\P\otimes S
\rightarrow E\otimes_{\At}\P\otimes S$, where $E$ is a
finitely
generated
projective module over $\At$ and $\P$ is a finitely generated
projective
module over $\At\times \Aht$. First,
we can reduce the problem to the case when
$E$ is a free $\At$ module. Indeed, since $E$ is a
finitely generated projective
$\At$ module it is a direct summand in $\At^k$ for some
$k$. We have:
$\At^k=E\oplus \tilde{E}$. Let $P_E$ be an
orthogonal projection on $E$. Choose any $\At$ connection
on $\tilde{E}$. Consider the Dirac operator $\hat{D}:
(E\oplus\tilde{E})\otimes_{\At}\P\otimes S\rightarrow
(E\oplus\tilde{E})\otimes_{\At}\P\otimes S$. Notice that
$\hat{D}=D\oplus \tilde{D}$, where $\tilde{D}$ is the
Dirac operator
on
$\tilde{E}\otimes_{\At}\P\otimes S$. Moreover, $P_E
\tilde{D}=\tilde{D}P_E=0$.
Therefore, if $\hat{Q}$ is a parametrix to  $\hat{D}$
then it is easy
to see
that $Q=P_E \hat{Q} P_E$ is a parametrix to $D$. More
precisely, if
$\hat{Q}\hat{D}=1+\hat{K}_1$ and
$\hat{D}\hat{Q}=1+\hat{K}_2$, then
$QD=1+K_1$ and $DQ=1+K_2$, where $K_1=P_E\hat{K}_1 P_E$ and
$K_2=P_E\hat{K}_2 P_E$. Notice that $K_1$ and $K_2$
preserve the
properties
of $\hat{K}_1$ and $\hat{K}_2$.

Similarly, we can assume that $\P$ is a free module over
$\At\times\Aht$.
Therefore, we need to construct parametrix only in the
case of Dirac
operator
$D$ on $(\At\times\Aht)^k\otimes S$. One can consider
$(\At\times\Aht)$
as the space of functions on the product of tori
$T^n\times T^n$.
The Dirac operator $D$ in this case becomes a sum of a usual
commutative
Dirac operator $D_c$ along the first torus plus a bounded
operator
$A$
(preserving the space of smooth functions),
that is $D=D_c+A$. Let $Q_c$ be a parametrix to $D_c$. Then,
$Q_c D=Q_c D_c+Q_c A$. It follows immediately from the
properties of
$Q_c$ that $Q_c A$ is bounded operator from $H_{(0)}$ to
$H_{(1)}$,
where $H_{(l)}$ are Sobolev spaces of functions along the
first
torus.
Thus, for any given natural number $l$ we can easily
construct an operator $Q$ (maps $H_{(0)}$ to $H_{(1)}$)
such that $QD=1+K_1$ and
$K_1$ maps $H_{(0)}$ to $H_{(l)}$. Similarly, for any
given natural
number $l$
we can construct an operator $Q$ (maps $H_{(0)}$ to $H_{(1)}$)
such that $DQ=1+K_2$ and $K_2$ maps $H_{(0)}$ to $H_{(l)}$.
Moreover, one can see that $Q$ can be chosen to be the
same. Such an
operator
$Q$ is enough for all our purposes. We call it parametrix
here.

Second, let us prove the following lemma.
\begin{lem}
\label{lemA1}
Let $P$ be an operator on $L^2(\R)\times \Aht$ commuting
with the
action
of $\Aht$. Assume that
$P^*=P$, $P^2=P$ and $P$ is ``compact'' ({\it i.e.},
$P\in {\cal K}
\hat{\otimes}\Aht$, where $\cal K$ is the algebra of compact
operators
on  $L^2(\R)$).
Then $P$ is a projection on a finitely
generated projective $\Aht$ module.
\end{lem}
{\bf Proof:} First, we can approximate $P$ by a self-adjoint
operator $A=\sum_j \langle f_j,\cdot\rangle g_j\times
a_j$, where
$f_j,g_j\in L^2(\R)$ and $a_j\in\Aht$ (since $P$ is
self-adjoint).
 That is, $P=A+\epsilon$, where
$\epsilon$ is a self-adjoint
operator of norm less then $\frac{1}{100}$. It is obvious
that $P$ is a projection on the kernel of $1-P$.
We can write
$1-P=1-\epsilon-A=(1-A(1-\epsilon)^{-1})(1-\epsilon)$.
If $(1-P)x=0$ it means that
$(1-A(1-\epsilon)^{-1})(1-\epsilon)x=0$.
Denote by $y=(1-\epsilon)x$. Then $(1-A(1-\epsilon)^{-1})y=0$.
{}From this we see that $y=\sum_j \langle
f_j,(1-\epsilon)^{-1}y\rangle
g_j\times a_j$. In particular this means that $y\in \sum_j
g_j\times\Aht$.
The module
$\sum_j g_j\times\Aht$ is a free finitely generated
$\Aht$ submodule
of
$L^2(\R)\times \Aht$. We denote it by $U$. Denote by $M$
the set
$(1-\epsilon)^{-1}U$. Since $(1-\epsilon)^{-1}$ is
invertible operator commuting with $\Aht$ we see that $M$
is $\Aht$
module.
It is not hard to check
(using the fact that $(1-\epsilon)^{-1}$ is self-adjoint
and close to
identity) that
$M$ is a free hermitian finitely generated $\Aht$ module.

{}From the above it is obvious that if $(1-P)x=0$ then
$x\in M$.
Therefore,
we can restrict projection $P$ on $M$ and we will get the
same image.
What is left to check is that  $P|_M$  is a projection
operator.
We see that if  $m\in M$ then $(1-P)Pm=
P m-P^2 m=Pm-Pm=0$. This implies that $P|_M$ is
projection operator.
Now, since $M$ is finitely generated free $\Aht$ module
and $P:M\rightarrow M$ is a projection lemma is proved. $\Box$

We cannot directly apply the above lemma to the module
$E\otimes_{\At}\P\otimes S$. But since
$E\otimes_{\At}\P\otimes S$
can be
considered a direct summand in $(\At\times\Aht)^k\otimes S$
apply Lemma \ref{lemA1} to $(\At\times\Aht)^k\otimes S$.

\section{Acknowledgments.}
\label{last}

We are  grateful to G.~Elliott, D.~Fuchs, A.~Gorokhovsky,
G. Kuperberg, A.~Lawrence, A.~Losev, M.~Rieffel,
S.~Shatashvili and
I.~Zakharevich for
 discussions. The research
of N.~N. was supported by Harvard Society of Fellows,
partially by
NSF
under grant PHY-98-02-709, partially by RFFI under grant
98-01-00327
and
partially by grant 96-15-96455 for scientific schools.
The research of A.~S. was partially supported by NSF under
grant
DMS-9500704.

\end{document}